\documentclass{PoS}

\title{The present status of the EPS nuclear PDFs}

\ShortTitle{The present status of the EPS nuclear PDFs}

\author{\speaker{Hannu Paukkunen}$^a$, Carlos A. Salgado$^{a}$
        and Kari J. Eskola$^{b}$ \\
\llap{$^a$}  University of Santiago de Compostela \\
             Departamento de F\'\i sica de Part\'\i culas and IGFAE, Universidade de Santiago de 
Compostela, Spain \\
\llap{$^b$}  University of Jyv\"askyl\"a \\
             P.O. Box 35, FI-40014 University of Jyv\"askyl\"a, Finland \\
Email: \email{hannu.paukkunen@usc.es}, \email{carlos.salgado@usc.es}, \email{kari.eskola@phys.jyu.fi}
}

\abstract{The recent global analyses of the nuclear parton distribution
functions (nPDFs) lend support to the validity of the factorization
theorem of QCD in high-energy processes involving bound nucleons.
With a special attention on our latest global analysis EPS09,
we review the recent developements in the domain of nuclear PDFs.}

\FullConference{35th International Conference of High Energy Physics\\
                 July 22-28, 2010\\
                 Paris, France}

\begin{document}

\section{Introduction}

\vspace{-0.3cm}
The global analyses of nPDFs approach the experimentally observed differences
between the hard-process cross-sections involving bound and free nucleons, $\sigma^{\rm bound \, nucleon} \neq \sigma^{\rm free \, nucleon}$, with the toolkit of perturbative QCD (pQCD).
Different processes in different kinematical regions are studied in order to
extract the universal --- process independent --- nuclear PDFs, and to 
estimate the conditions in which the factorization theorem of QCD appears applicable.
In this talk we recapitulate the essentials of the latest offspring of
the EPS-generation \cite{Eskola:1998iy,Eskola:1998df,Eskola:2007my,Eskola:2008ca},
EPS09 \cite{Eskola:2009uj}, and release a piece of news about the recent
EPS-activities.

\vspace{-0.3cm}
\section{Analysis}

\vspace{-0.3cm}
\subsection{Experimental Input}

The main source of experimental input in EPS09 comes from the measurements of
deep inelastic scattering (DIS) and Drell-Yan (DY) dilepton production. These
data alone, however, leave the gluon PDFs badly unconstrained. As noted in
\cite{Eskola:2009uj}, part of such freedom can be reduced --- in a way that do
not cause disagreement with other data --- by adding data for average high-$p_T$ pion 
production in d+Au collisions measured at RHIC. It should be noted that the
fragmentation functions which in pQCD-calculations turn the partons to pions
might also experience a modification with respect to the free proton
fragmentation functions \cite{Sassot:2009sh}. This effect is not accounted
for in EPS09 where the pion-observables are mainly needed for constraining the
nuclear gluons.


\vspace{-0.3cm}
\subsection{Theoretical Framework}

In EPS09, the bound proton PDFs for each flavor $i$ are linked to the 
free proton ones by
\begin{equation}
 f_i^{\rm bound \, proton}(x,Q^2) \equiv R_i^{A}(x,Q^2) f_i^{\rm free \, proton}(x,Q^2).
\end{equation}
The nuclear modification factors $R_i^{A}$ are parametrized at charm quark mass treshold
$Q^2 = Q^2_0 = 1.69 \, {\rm GeV}^2$, assuming a flavor-independent nuclear modification
for light sea quarks $R^{A}_{\overline{u}} = R^{A}_{\overline{d}} = R^{A}_{\overline{s}}$,
and valence quarks $R^{A}_{u_V} = R^{A}_{d_V}$ at $Q^2_0$. This is, of course, just a simplifying
assumption and our plan is to to release this assumption in a future.


The EPS09 employ the $\overline{MS}$, zero-mass variable flavor number scheme.
Accordingly, the baseline free proton PDF set used was the CTEQ6.1M \cite{Stump:2003yu},
which probably remains as the last CTEQ-set in this simple scheme, the more recent ones involving
a more complete prescription to treat the heavy quarks. Recently, we have also implemented
such a scheme to our fitting engine. However, the available nuclear data
are not very sensitive to the heavy quarks, and our tentative results do not display
large deviations from the zero-mass ones. Therefore, the usage of EPS09 with newer sets of the free
proton PDFs like CT10 \cite{Lai:2010vv} should be justified.

\vspace{-0.3cm}
\subsection{Fit Quality}

In order to find the $R_i^{A}$s that optimally reproduce the experimental data, a certain 
quality criterion is need. We minimize a generalized $\chi^2$-function which is essentially
\begin{equation}
 \chi^2 \equiv \sum_N w_N \sum_{i \in N} \left(\frac{D_i-T_i}{\sigma_i}\right)^2,
\end{equation}
where $D_i$ and $\sigma_i$ are the experimental data points with their errors, and $T_i$
is the corresponding calculated value. A small extension to this due to the additional normalization
uncertainty of the pion data is explained in \cite{Eskola:2009uj}. The additional weights $w_N$ are assigned
to emphasize certain data sets with only a small amount of data points but carrying physically
relevant information. Partly, this is a technical issue aiming to improve the convergence of the fit with more open
parameters, but certainly introduce some subjectivity to the central fit. Unfortunately,
the weights also distort the error analysis explained shortly, and inevitably lead to a
somewhat underestimated PDF-errors. Reducing the impact of such reweighting is one of
the issues we aim to improve in the future.

\vspace{-0.3cm}
\subsection{Error Analysis}

The uncertainty analysis propagates the experimental errors from the data to the PDFs.
The Hessian method, used in EPS09, relies on a quadratic approximation
of $\chi^2$ around its minimum $\chi^2_0$
\begin{equation}
 \chi^2(\{ a_i\}) \approx \chi^2_0 + \sum_{ij} \delta a_i H_{ij} \delta a_j = \chi^2_0 + \sum_i z_i^2,
\end{equation}
where the $\delta a_i$s are the deviations of the original fit parameters $a_i$ from the
best-fit values. The key point is that the Hessian matrix $H_{ij}$ can be diagonalized
thereby finding an uncorrelated basis $\{ z_i \}$. Relying to so called 90\% confidence
criterion, we find that that distorting the best fit in the directions of each $z_i$ such
that the $\chi^2$ changes by 50, still gives a resonable agreement with the data. These corners
of parameters define the uncertainty sets $S_i^\pm$ facilitiating the estimation of the upper$\backslash$lower
error in any PDF-dependent quantity X with best-fit prediction $X(S_0)$, by
\begin{eqnarray}
\Delta X^{+ \backslash -} & = & \sqrt{\sum_i \max \backslash \min \left[ X(S_i^+) - X(S_0), X(S_i^-) - X(S_0),0 \right]^2}. \end{eqnarray}

\vspace{-0.3cm}
\subsection{Results}

\begin{figure}[!htb]
\center
\vspace{-1.0cm}
\includegraphics[scale=0.55]{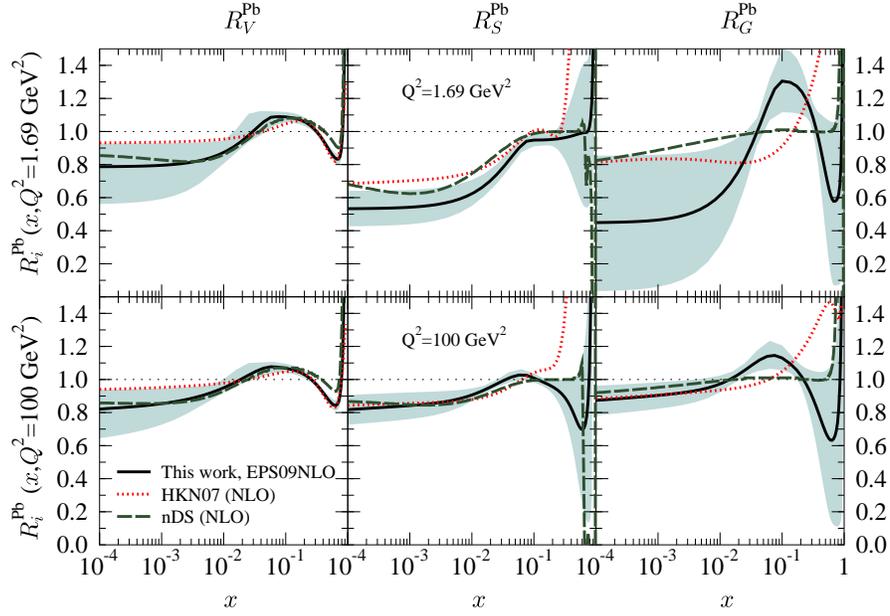}
\caption[]{\small The nuclear effects for valence quarks $R_V$, sea quarks $R_S$, and gluons
$R_G$ in Lead. The NLO results from EPS09, HKN07 \cite{Hirai:2007sx}, and nDS \cite{deFlorian:2003qf} 
are shown at two values of $Q^2$.}
\label{Fig:Modif}
\end{figure}

We summarize the resulting nuclear effects in PDFs by Figure~\ref{Fig:Modif} which shows
the nuclear effects in Lead at a low scale $Q^2 = 1.69 \, {\rm GeV}^2$ and also at a higher scale
$Q^2 = 100 \, {\rm GeV}^2$ to demonstrate the effect of DGLAP evolution. For lighter nuclei
the effects are closer to unity. The blue bands denote the EPS09-uncertainty.
We also show the results from two alternative analyses \cite{Hirai:2007sx,deFlorian:2003qf} helping
to appreciate the well and badly constrained components of the nuclear effects.

\vspace{-0.3cm}
\section{Recent Activities}

\vspace{-0.3cm}
For studying the universality of nPDFs further, data from other processes
are always welcome. One of such processes is the neutrino beam induced DIS.
It has been suggested \cite{Schienbein:2007fs} that the NuTeV $\nu{\rm Fe}$-data 
\cite{Tzanov:2005kr} favors different nuclear effects than extracted from charged
lepton DIS and DY processes. However, our analysis \cite{Paukkunen:2010hb} with a larger
set of data implies that this is a sheer artefact brought about looking only
at the NuTeV data that appear to contain internal incosistencies.

In order to cleanly probe the nPDFs, an electron-ion collider like the planned LHeC
would be an ideal machine. By fitting to a simulated LHeC data, we have found that
such data would dramatically improve the determination of small-$x$ gluons and sea
quarks. Another promising probe for nPDFs, that could be available before the realization of
an electron-ion collider, would be the proton-ion runs at LHC. We have
found that e.g. production of heavy electroweak bosons in such a collisions would
provide a surprisingly neat tool to investigate the nuclear modifications to the PDFs
with not much sensitivity to the underlying set of free proton PDFs. Both results
mentioned here will be published shortly.

\end{document}